\documentclass[a4paper,11pt]{article}
\usepackage{pos}
\usepackage{graphicx}
\usepackage{subfig}
\usepackage{siunitx}

\title{Statistical analysis and correction of the pile-up effect in MAPMT single photoelectron counting with the SPACIROC-3 ASIC: application to the Mini-EUSO experiment}
\ShortTitle{Statistical analysis and correction of the pile-up effect in extended dead-time electronics}
 
\author[a]{Enzio M'sihid*}
\author[a]{Etienne Parizot}
\author[b]{Matteo Battisti}
\author[a]{Sylvie Blin}
\onbehalf{for the JEM-EUSO Collaboration\\[-1mm]{\normalsize \normalfont (a complete list of authors can be found at the end of the proceedings)}}

\affiliation[a]{Université Paris Cité, CNRS, Laboratoire AstroParticule et Cosmologie, \\ 10 Rue Alice Domon et Léonie Duquet, 75013, Paris, France}
\affiliation[b]{Instituto Nazionale di Fisica Nucleare, Università degli Studi di Roma Tor Vergata,\\
Via della Ricerca Scientifica, 00133 Roma RM, Italy}

\emailAdd{msihid@apc.in2p3.fr}
\emailAdd{parizot@apc.in2p3.fr}
\emailAdd{matteo.battisti@edu.unito.it}

\abstract{We present a comprehensive study addressing pile-up effects in single photoelectron counting with R-11265 Hamamatsu multi-anode photomultiplier tubes (MAPMTs) equipped with the SPACIROC-3 ASIC. Extended dead time in the electronics causes saturation and quenching of the counting rate, an effect we counter by inverting the pile-up plot once the double pulse resolution is determined. Our work combines extensive numerical simulations with experimental validations to quantify the statistical uncertainties associated with the corrected event rates. We apply this methodology to the Mini-EUSO experiment onboard the International Space Station where machine learning techniques are employed to extract pixel-by-pixel double pulse resolutions from long-term photon count histograms. This integrated approach enables the accurate recovery of true photon fluxes essential for studying ELVES, meteors and other transient phenomena detected by Mini-EUSO.}

\ConferenceLogo{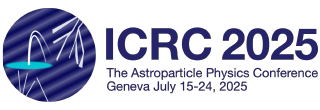}

\FullConference{39th International Cosmic Ray Conference (ICRC2025)\\
 15–24 July 2025\\
Geneva, Switzerland\\}

\begin{document}
\maketitle

\section{Introduction and scientific context}
Accurate photon counting is essential in astroparticle experiments employing multi-anode photomultiplier tubes (MAPMTs) in single photoelectron mode, where event reconstruction relies on precise photon rate measurements. However, the presence of electronic dead time in the readout introduces pile-up: successive photons within short intervals lead to under-counting, impairing detector linearity and reconstruction accuracy.

We present a correction framework combining numerical simulations, analytical modeling from renewal theory, and experimental validation to recover true event fluxes from an observed number of count. Applied to Mini-EUSO, a UV telescope on the International Space Station ~\cite{Mini-EUSO} equipped with SPACIROC-3 ASICs~\cite{ASIC}, this method enables pixel-by-pixel estimation of effective dead time and systematic pile-up correction. These results enhance event counting reliability for current and future missions utilizing high-speed detectors.

\section{Modeling and characterization of pile-up effects}
\label{sec:pileupcharacterization}
This section presents a numerical and theoretical framework for modeling extended dead-time effects, combining Monte Carlo simulations and renewal theory to characterize saturation behavior. The model predictions are validated against experimental data under controlled increasing-rate conditions on an MAPMT equipped with the SPACIROC-3 ASIC.

\subsection{Numerical simulation and method for uncertainty estimation}

To quantify the effect of pile-up on the mean counting rate and its fluctuations, we implemented a numerical simulation that models stochastic photon arrivals and their registration by a detector with finite double pulse resolution, $\tau$. It provides a statistical mapping between the true event rate $\rho$ and the observed counts $N_\mathrm{count}$ over an integration window of fixed duration, $T$, enabling the estimation of the variance of each variable, when the other is fixed.

\begin{figure}[t]
\centering
\hfill
\includegraphics[width=0.47\textwidth]{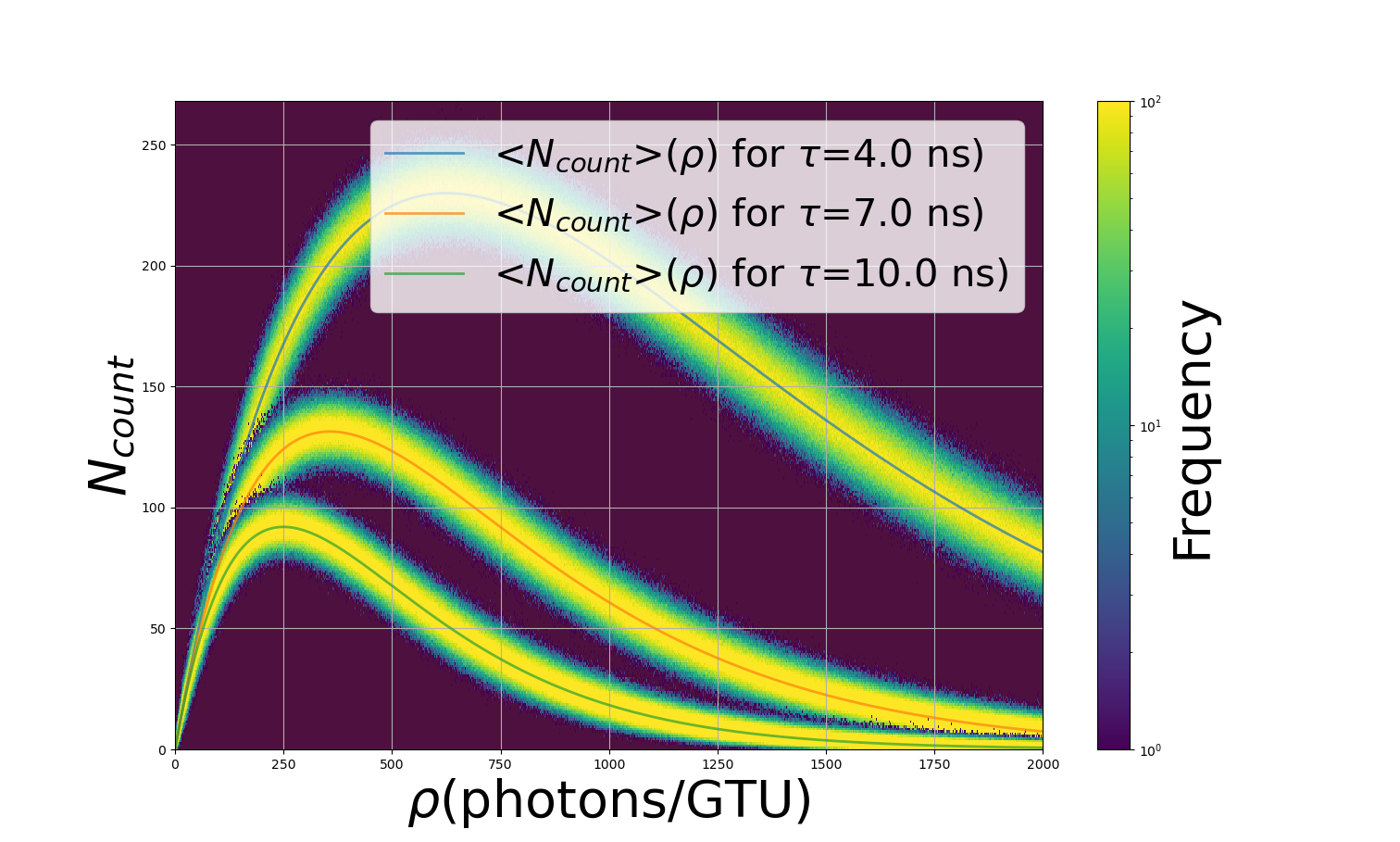}
\hfill
\includegraphics[width=0.35\textwidth]{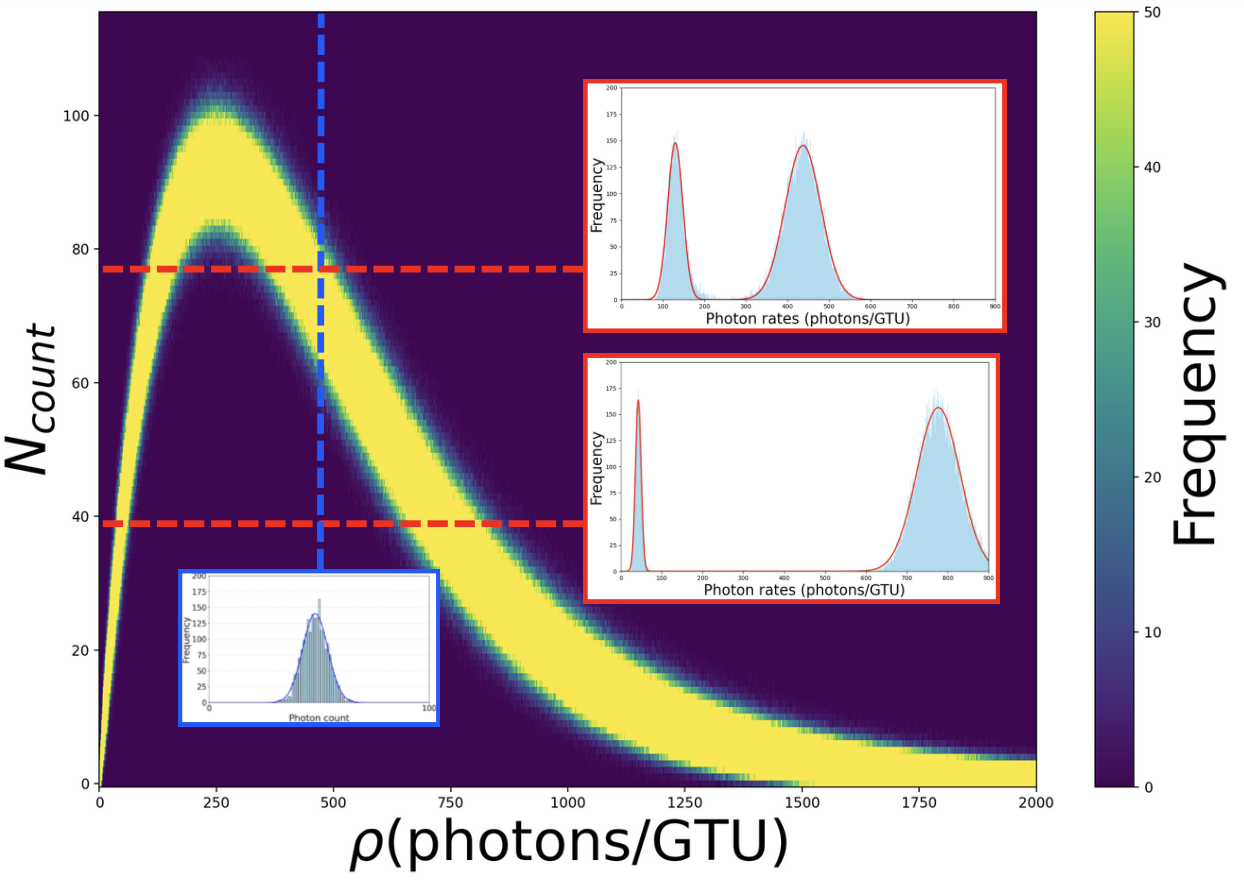}
\hfill
\vspace{-0.5em}
\caption{Left: pile-up plot showing the number of occurrences (color code) of a given photon count (in ordinate) as a function of the photon rate (in abscissa), for a three different values of the double pulse resolution $\tau$. The solid lines show the average $\langle N_\mathrm{count}\rangle(\rho)$, also known as the pile-up curve or saturation curve. Right: example of histograms obtained from vertical (blue) and horizontal (red) cuts of a pile-up plot corresponding to $\tau = 10$~ns and $T = 2500$~ns.}
\label{fig:saturation-curve}
\end{figure}

Photon arrivals are modeled as a Poisson process of rate $\rho$ over a time interval $T$ (so-called gate time unit or GTU, typically 2500 ns), with pile-up simulated by rejecting events within a dead time $\tau$ after each event. This mimics an extended dead-time behavior expected for SPACIROC-3 electronics. The simulation is executed over a wide range of values of the rate $\rho$ (0–2000~ph/GTU), and repeated for $N_\mathrm{GTU}$ independent GTUs for each rate. For each GTU, $i \in [1,N_\mathrm{GTU}]$, we count the number of effectively detected photons, $N_\mathrm{count}(i;\rho)$, at the current rate. This provides a count distribution as shown on the left panel of Fig.~\ref{fig:saturation-curve} for three different values of $\tau$. The color shows the number of times a given number of photons (in ordinate) has been counted for a given photon rate (in abscissa). The solid lines show the average counts, $\langle N_\mathrm{count}\rangle$, as a function of $\rho$. As can be seen, $\langle N_\mathrm{count}\rangle$ first increases almost linearly with $\rho$ when the photon rate is low enough for the probability of pile-up to be negligible. It then reaches a maximum at a value of $\rho$ directly related to the dead time $\tau$ (see below), and finally decreases down to fewer and fewer counts, when the probability for the interval between two consecutive events to be larger than $\tau$ becomes smaller and smaller.

To estimate the uncertainties associated with $N_\mathrm{count}$ (at fixed $\rho$) and $\rho$ (at fixed $N_\mathrm{count}$), we analyze the histograms obtained from the simulation by performing cuts of the pile-up plot along vertical and horizontal lines. Examples of such cuts are shown on the right panel of Fig.~\ref{fig:saturation-curve}. Simple gaussian (or double gaussian) fits then give access to the averages and dispersions of the counts at a given rate, $\langle N_\mathrm{count}\rangle(\rho)$ and $\sigma_\mathrm{count}(\rho)$, as well as the average and dispersion of rates for a given count, $\langle\rho\rangle(N_\mathrm{count})$, and $\sigma_{\rho}(N_\mathrm{count})$.

\subsection{Extended dead-time modeling within the renewal theory framework}

The theoretical framework that allows to derive analytically $\langle N_\mathrm{count}\rangle(\rho)$, $\sigma_\mathrm{count}(\rho)$, $\langle\rho\rangle(N_\mathrm{count})$ and $\sigma_{\rho}(N_\mathrm{count})$ in such a context is the so called renewal theory~\cite{muller1971}. Within this framework, the expected number of counts $\langle N_\mathrm{count}\rangle(\rho)$ observed over a fixed time interval $T$ for a constant dead time $\tau$ is given by ~\cite{muller1971, pomme2015} :

\vspace{-0.4em}
\begin{equation}
\langle N_\mathrm{count}\rangle(\rho) = \rho T e^{-\rho \tau}.
\label{eq:N_expected}
\end{equation}

The inversion of this expression (to infer $\rho$ from $N_\mathrm{count}$) uses the Lambert W function implicitly defined by $z = w e^{w} \Longleftrightarrow\; w = W(z)$:

\vspace{-0.4em}
\begin{equation}
\langle\rho\rangle(N_\mathrm{count}) = -\frac{1}{\tau} W\left(-\frac{\tau N_\mathrm{count}}{T}\right).
\label{eq:rho_from_N}
\end{equation}

Furthermore, the variance of the observed number of counts for a given event rate is found to be lower than the variance of a Poissonian process with the same average, and can be expressed as ~\cite{muller1977,pomme2015} :

\vspace{-0.4em}
\begin{equation}
\sigma_\mathrm{count}(\rho)^2 = N_\mathrm{count} \left(1 - 2 \rho \tau e^{-\rho \tau}\right).
\label{eq:sigma_N}
\end{equation}

Propagating the uncertainty on $\tau$ (a parameter to be determined experimentally) as well as the dispersion $\sigma_\mathrm{count}$ for a given $N_\mathrm{count}$ measured in one particular GTU, we obtained the associated uncertainty on the true event rate $\rho$ as:

\begin{equation}
\begin{split}
    \sigma_\rho(N_{\mathrm{count}})^{2} &= \left| \frac{d\rho}{dN_{\mathrm{count}}} \right|^2 \sigma_\mathrm{count}(\rho)^2 + \left| \frac{d\rho}{d\tau} \right|^2 \Delta\tau^2 \\
 & = 
\frac{W\left(-\frac{\tau N_\mathrm{count}}{T}\right)^2}{\tau^2}
\left[ \left(1 - \frac{1}{1 + W\left(-\frac{\tau N_\mathrm{count}}{T}\right)} \right)^2 \frac{\Delta \tau^2}{\tau^2} 
+ \left( \frac{1}{1 + W\left(-\frac{\tau N_\mathrm{count}}{T}\right)} \right)^2 \frac{\sigma_{\mathrm{count}}(\rho)^2}{N_\mathrm{count}^2} 
\right].
    \end{split}
\label{eq:sigma_rho}
\end{equation}

These analytical results allow us to correct for pile-up effects, not only to infer the most probable event rate associated with a recorded number of counts, but also its uncertainty. It can be used for any event counting set-up operating with some extended dead time $\tau$. In the case when different events are associated with different values of $\tau$ (without correlation between them), the same formula apply with $\tau$ replaced by the statistical average of the different values.

\subsection{Experimental characterization of the pile-up phenomenon in an MAPMT coupled with a SPACIROC-3 ASIC}

To validate the theoretical and numerical predictions, we conducted laboratory measurements using a Multi-Anode Photomultiplier Tube (MAPMT) of type R-11265 from Hamamatsu, coupled to a SPACIROC-3 ASIC in single photoelectron counting mode. This campaign aimed to characterize the detector’s response under increasing light levels to observe the onset of pile-up, and confirm the validity of the extended dead-time assumption.

\subsubsection{Experimental set-up}

The measurements were conducted using a “lightscan” procedure, where a continuous 404~nm LED illuminates one pixel of an MAPMT inside a dark box (see \cite{callibration2} for details). The light intensity is progressively increased across acquisition steps by adjusting the LED driving current and recorded by a calibrated low gain photodiode (LGPD) connected to a powermeter. For each intensity, the number of photons actually counted by the corresponding ASIC channel are recorded for each of a total of 10,000 GTUs of 2500~ns.

This procedure allows us to explore the detector's response from the linear regime into the pile-up-dominated region, while providing counting statistic to extract $\langle N_\mathrm{count}\rangle(\rho)$ and $\sigma_\mathrm{count}(\rho)$.

\subsubsection{Lightscan outputs and comparison with theory}

\begin{figure}[t]
\centering
\hfill
\includegraphics[width=0.48\textwidth]{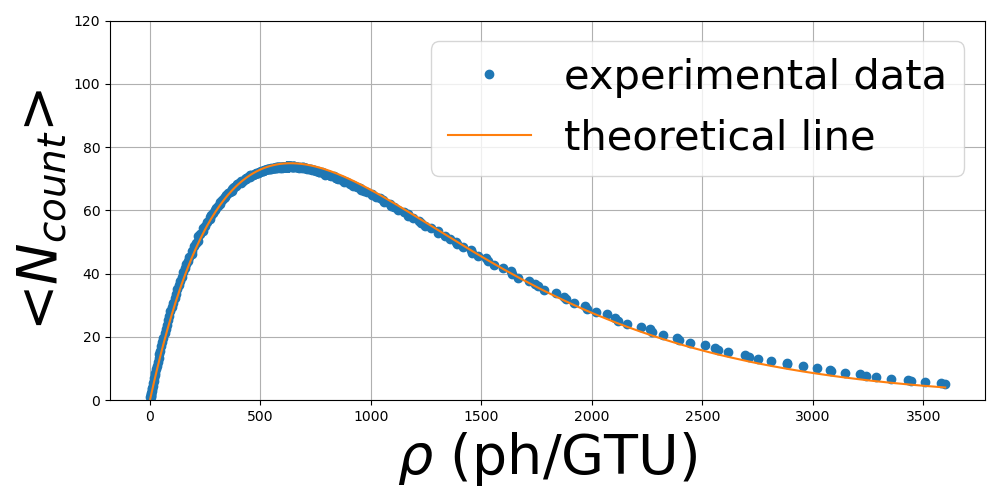}
\hfill
\includegraphics[width=0.48\textwidth]{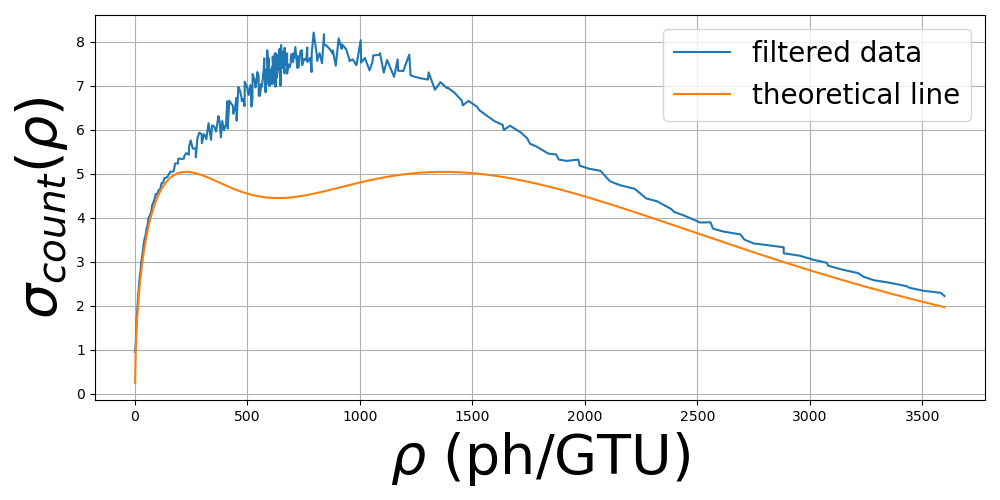}
\hfill
\vspace{-0.5em}
\caption{Left : Fit of the experimental $\langle N_\mathrm{count}\rangle(\rho)$ with eq.(\ref{eq:N_expected}), multiplying $\rho$ by a quantum efficiency $\epsilon_{\mathrm{quantum}}$ (to convert electron into photoélectron), yielding  $\tau = 12.3$~ns and $\epsilon_\mathrm{quantum} = 0.32$. Right : comparison of the experimental $\sigma_\mathrm{count}(\rho)$  with its expected shape from eq.(\ref{eq:sigma_N}) for the fit parameters yielded by the fit of the average. Both experimental data were obtained from a lightscan on a single channel of an MAPMT connected to a SPACIROC3-ASIC working at a GTU of 2500~ns.}
\label{fig:sigmaNexp}
\end{figure}

The lightscan data was analyzed by converting the powermeter reading from the LGPD into its corresponding expected photons per GTU previously referred to as $\rho$ (see \cite{callibration1} for details). As shown in Fig.~\ref{fig:sigmaNexp}, left panel, the evolution of the average number of counts $\langle N_\mathrm{count}\rangle$ with $\rho$ follows the theoretical expectation (Eq.~(\ref{eq:N_expected})) with very high precision. However, the dispersion in the $N_\mathrm{count}$ values at fixed $\rho$ appears larger than expected for intermediate photon rates (around the maximum of the pile-up curve), as seen on the right panel of Fig.~\ref{fig:sigmaNexp}, where the red line shows the $\sigma_\mathrm{count}$ corresponding to the value of $\tau$ inferred from the fit in the left panel. On this part of the curve, the data also appear much noisier.

Upon closer inspection we identified that for high values of the light intensity, a few GTUs exhibit anomalous count values randomly distributed among the 10,000 GTU sequence, with spikes several sigmas away from the average. This may be related to disturbances in the ASIC baseline and/or threshold level, possibly due to large current drain to supply the electron multiplication process in the MAPMT. These outlier GTUs strongly distorts the calculated standard deviation (cf. blue curve in Fig.~\ref{fig:sigmaNexp}). This effect will be investigate in a future work. In the present work, we mitigated it at the analysis level by implementing a filtering procedure, applying a windowing around each $\langle N_\mathrm{count}\rangle(\rho)$ of ±3$\sigma_\mathrm{count}(\rho)$. GTUs falling outside this range were excluded from the calculations. A wider window of ±4$\sigma_\mathrm{count}(\rho)$ was also tested and led to very similar results, indicating that the affected GTUs are not consistent with statistical fluctuations expected from the pile-up model, but rather stem from systematic or electronic artifacts.

After the filtering, the experimental standard deviation aligns very  closely with the theoretical and simulated expectations for an extended dead time behavior as can be seen in Fig.~\ref{fig:overlap}, described in the next section.

\section{Results}
\label{sec:results}

\begin{figure}[h!]
\centering
\includegraphics[width=0.35\textwidth]{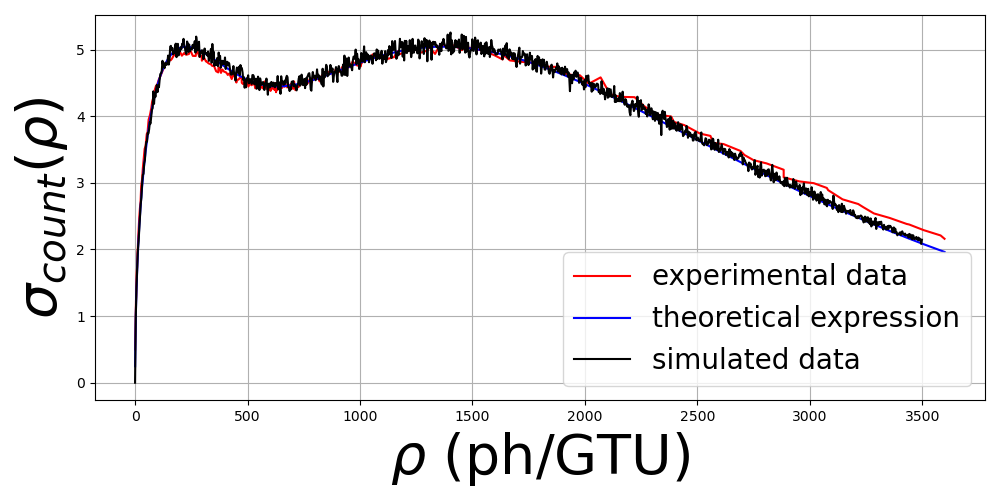}
\vspace{-0.5em}
\caption{Comparison of $\sigma_\mathrm{count}(\rho)$ obtained from renewal theory (blue line), Monte Carlo simulation (black line), and a lightscan (red line). Both the simulated and theoretical curves assume the same fixed dead time $\tau = 12.3 $~ns and a quantum efficiency of $\epsilon_{\mathrm{quantum}} = 0.32$, both obtained from the fit of the experimental data. All curves are drawn for a GTU of \SI{2500}{\nano\second}.}
\label{fig:overlap}
\end{figure}

The analytical modeling, numerical simulation, and experimental measurements described in Sect.~\ref{sec:pileupcharacterization} can be compared through their respective estimation of $\sigma_\mathrm{count}(\rho)$, as can be seen in Fig.~\ref{fig:overlap}. The curves show excellent agreement across the full dynamic range. A deviation observed in the experimental data above 1500ph/GTU is currently under investigation and may be linked to intrinsic limitations of the ASIC.

This overall consistency confirms that the extended dead-time model can accurately describe the SPACIROC-3 response. It also validates the renewal-theory formulation and confirms the Monte Carlo simulation’s accuracy in reproducing the statistical behavior of the single photon counting capability of the Mini-EUSO camera, establishing a solid foundation for its application to flight data correction.

\section{Pile-Up correction applied to Mini-EUSO flight data}
\label{sec:minieuso}

We apply the developed pile-up correction framework to flight data from the Mini-EUSO telescope. The objective is to extract the effective dead time $\tau$ for each pixel by fitting features of their count histograms with simulated ones. Once $\tau$ is known, observed counts can be inverted to recover the true photon flux, as explained above. This method is validated by the agreement between theory, simulation, and experimental data established in Section~\ref{sec:results}.

\subsection{Per-Pixel histogram analysis using supervised machine learning}
\label{subsec:ml_histogram}

\begin{figure}[htbp]
\centering
\hfill
\includegraphics[width=0.32\textwidth]{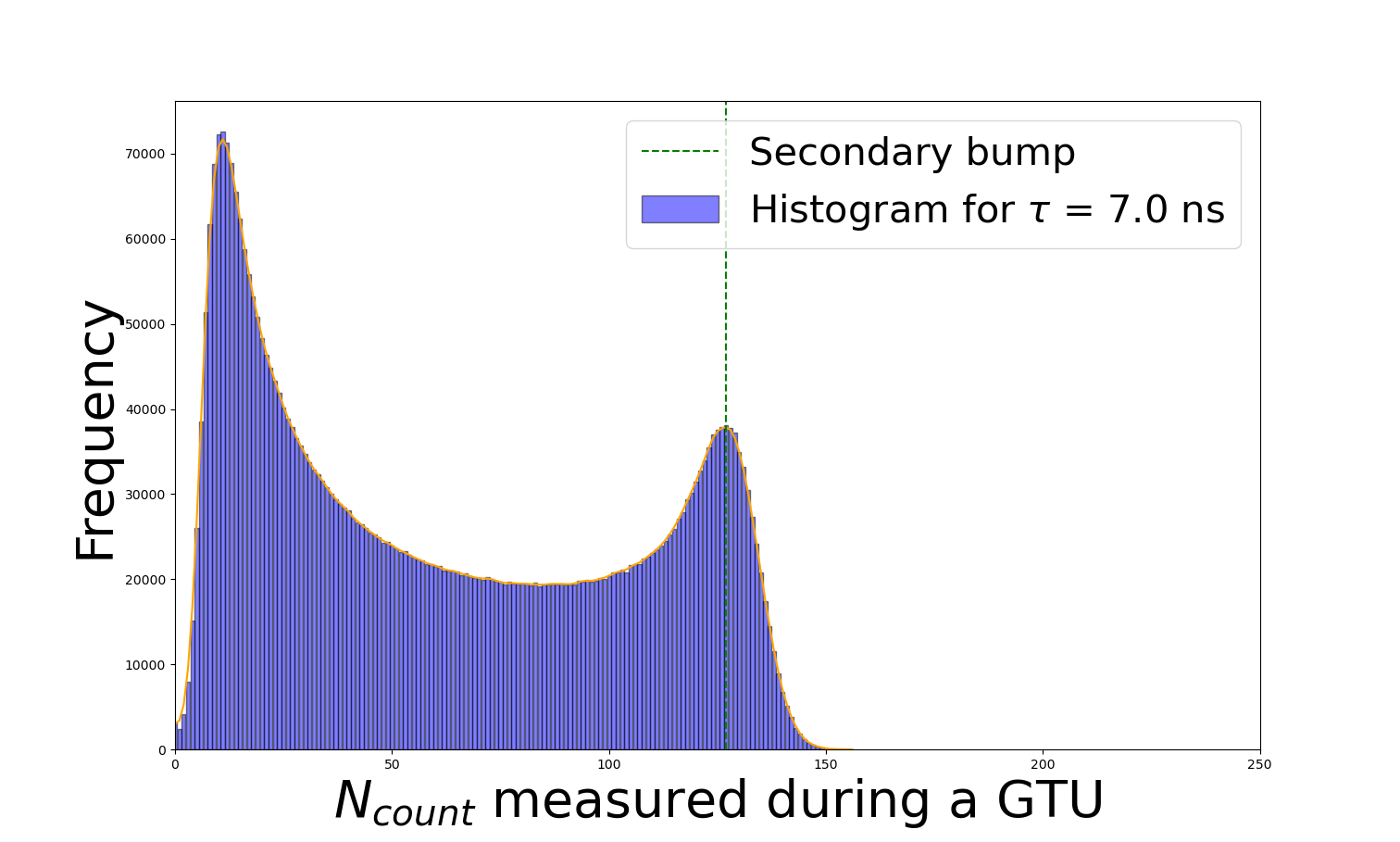}
\hfill
\includegraphics[width=0.32\textwidth]{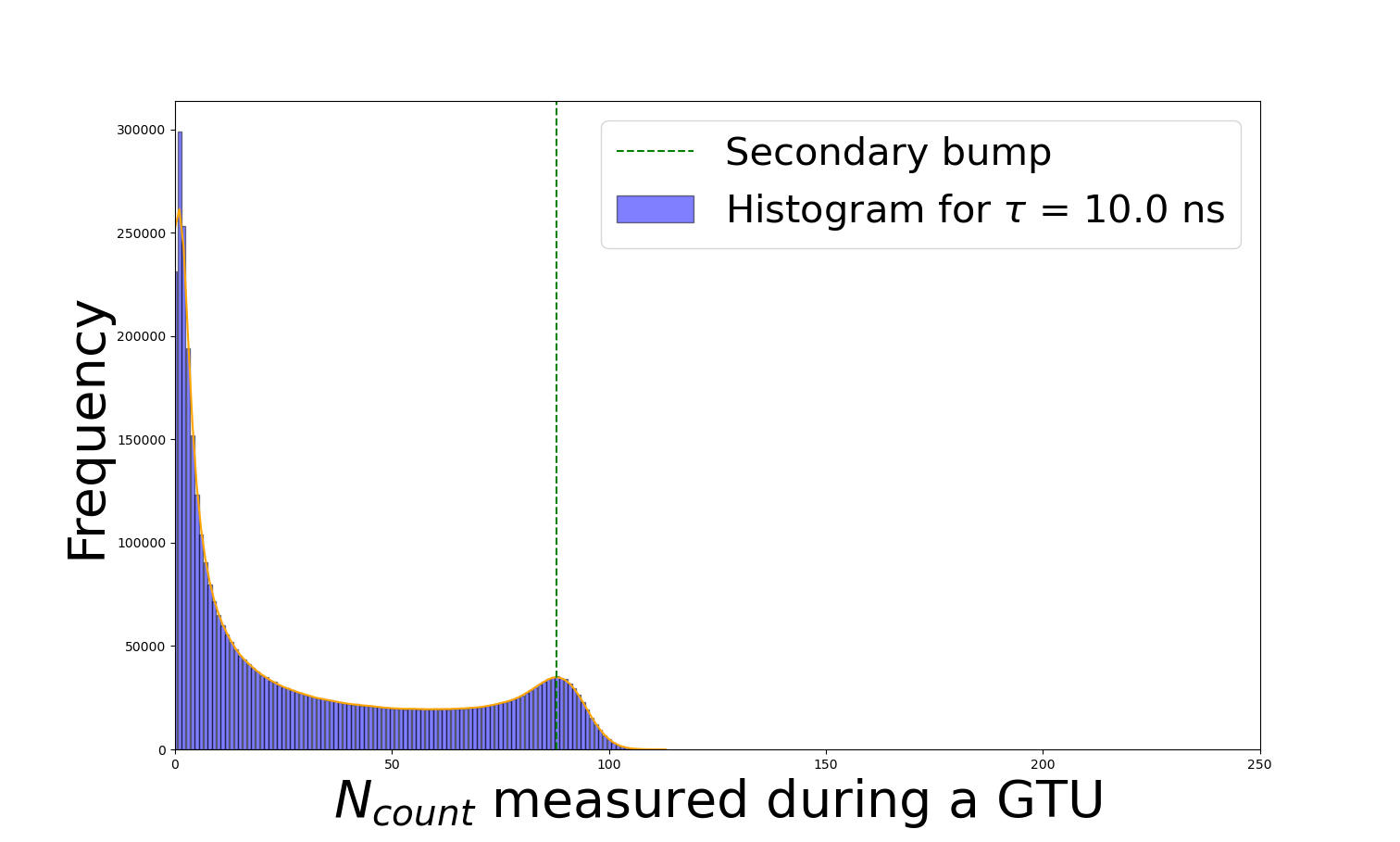}
\hfill
\includegraphics[width=0.32\textwidth]{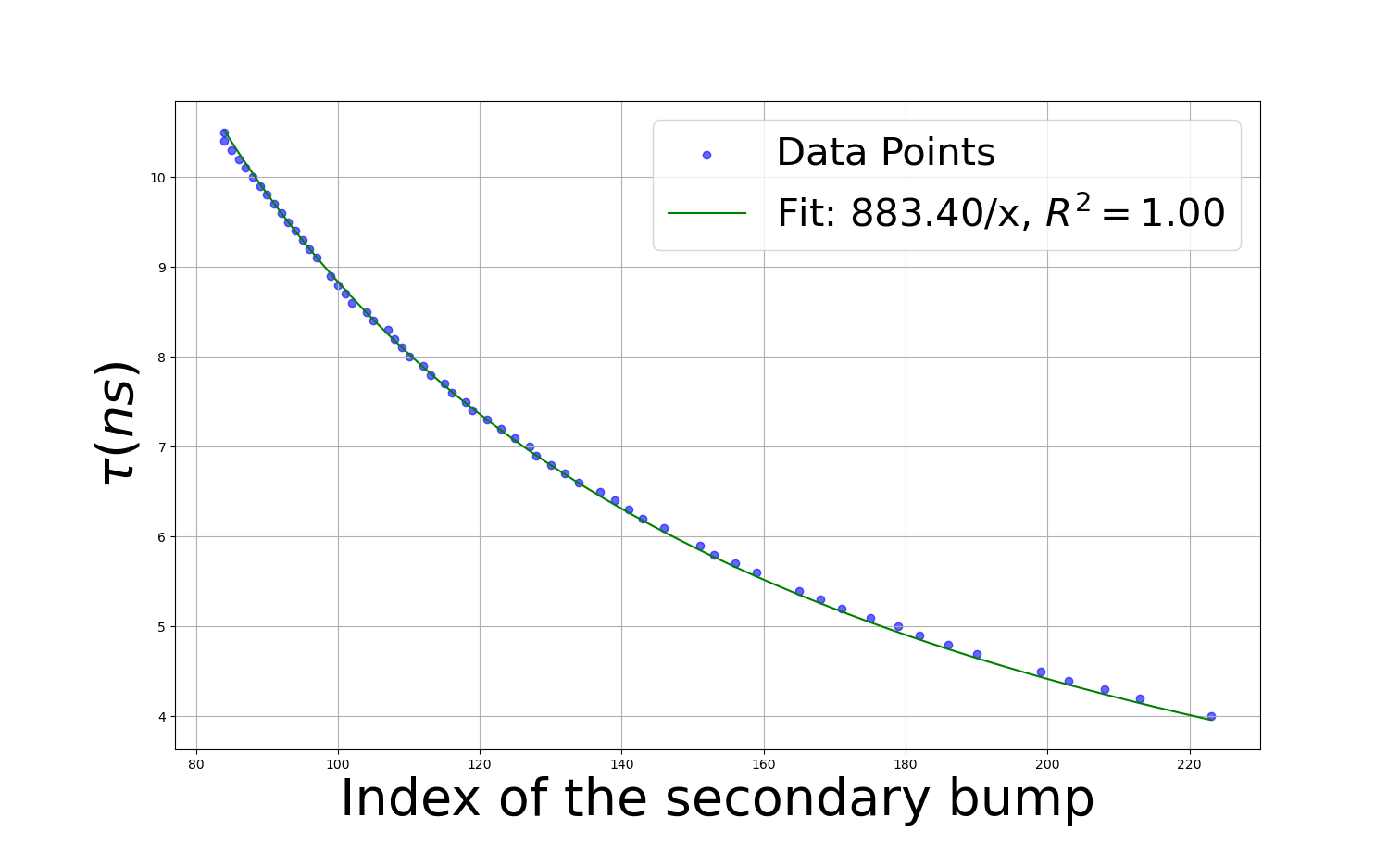}
\hfill
\vspace{-0.5em}
\caption{Left and Center : simulated count histograms showing the index of the secondary bump for two different $\tau$ and a GTU of \SI{2500}{\nano\second}. Right : Fitted dead time values as a function of the secondary bump index of the simulated histograms.}
\label{fig:dt_fit}
\end{figure}

The effective dead time $\tau$ of each pixel is estimated by analyzing long-term photon count histograms accumulated over several years. Pile-up-induced saturation feature – particularly the tail shape and the position of secondary structures such as the secondary bump – are sensitive indicators of the detector’s double pulse resolution. By comparing these features with a large dataset of simulated histograms generated for known $\tau$ values and uniform event rate distribution, using our Monte Carlo model, we establish a  correspondence between specific histogram signatures and the underlying dead time (see Fig.~\ref{fig:dt_fit}).

In the actual data registered by Mini-EUSO, the distribution of incoming photon rates is not uniform. However, the large variety of light intensities seen by the different pixels near their counting maximum allows to capture a similar behaviour, with a bump in the count histogram reminiscent of that obtained in the simulations (Fig.~\ref{fig:dt_fit}).

\begin{figure}[htbp]
\centering
\hfill
\includegraphics[width=0.48\textwidth]{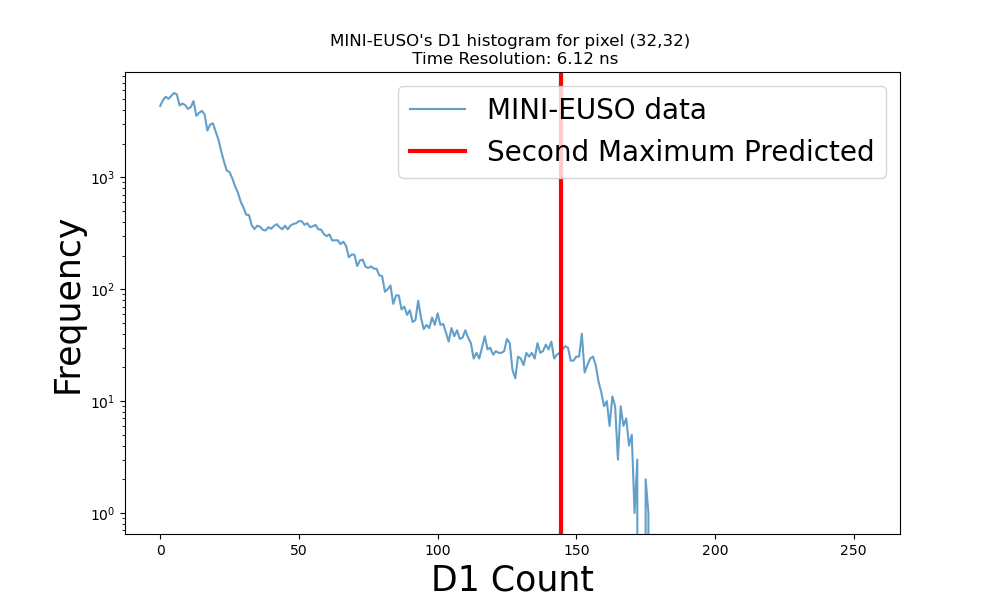}
\hfill
\includegraphics[width=0.48\textwidth]{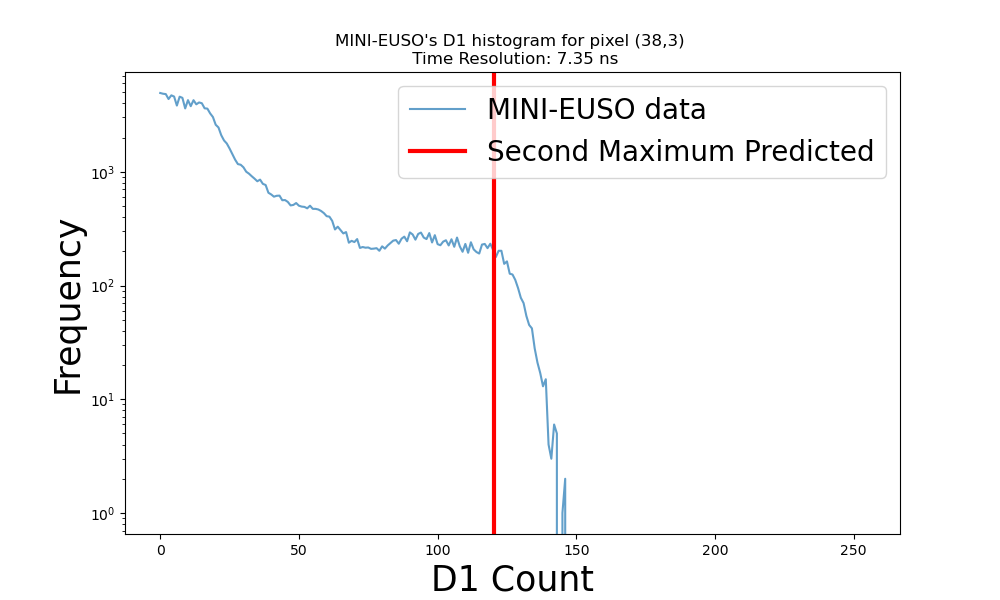}
\hfill
\vspace{-0.5em}
\caption{Prediction of the secondary bump index by the Random Forest Regressor model for two Mini-EUSO pixel's count histograms. The histograms are build from the data of sessions 20 to 44 and corresponds to the pixel (32,32) and (38,3) of the focal surface.}
\label{fig:minieusocounthist}
\end{figure}

To determine the value of $\tau$ for any given pixel of Mini-EUSO, we thus need to estimate the position of the secondary bump in its photon count histogram. Given the diversity of histogram shapes for the 2304 pixels, conventional algorithmic approaches proved unreliable, motivating the use of supervised machine learning. A training set was manually labeled from 150 representative pixels, and multiple models were tested. The best performance was achieved using Random Forest and XGBoost regressors, both reaching a mean absolute error of approximately 8.5 counts after hyperparameter optimization. Examples of bump position predictions are shown in Fig.~\ref{fig:minieusocounthist} for two different pixels.

The procedure was applied to all \( 48 \times 48 = 2304 \)  pixels of the Mini-EUSO focal surface. For each pixel, both machine learning models predicted the secondary bump index using 10 random seeds. Consistency across seeds within a model, and agreement between the two models, served as a validation criterion. Pixels exhibiting abnormal behavior – due to low efficiency or bit-shift artifacts (previously identified in the Mini-EUSO data analysis) – were flagged for exclusion or further inspection. Once the index is obtained, the corresponding dead time $\tau$ is inferred via the fitted calibration curve (see Fig.~\ref{fig:dt_fit}).

\subsection{Results of Mini-EUSO pile-up analysis}
\label{subsec:ml_results}

\begin{figure}[htbp]
\centering
\hfill
\includegraphics[width=0.48\textwidth]{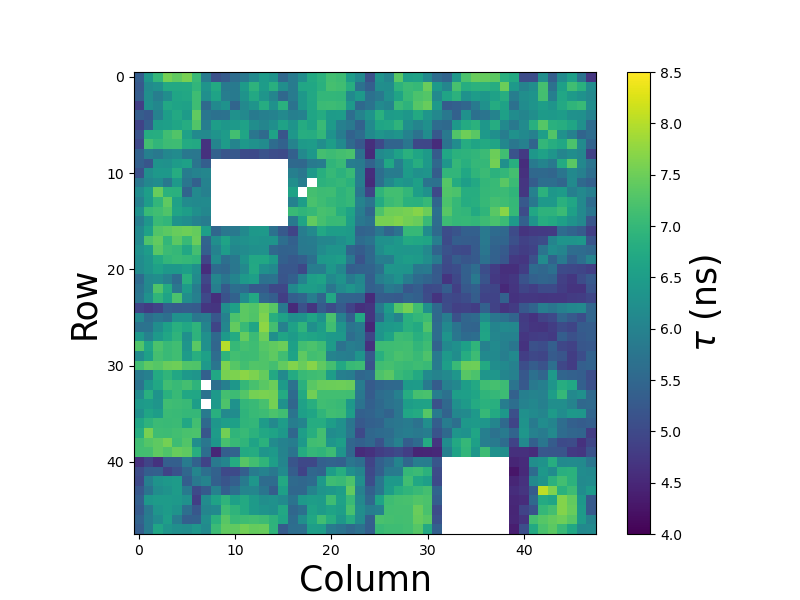}
\hfill
\includegraphics[width=0.48\textwidth]{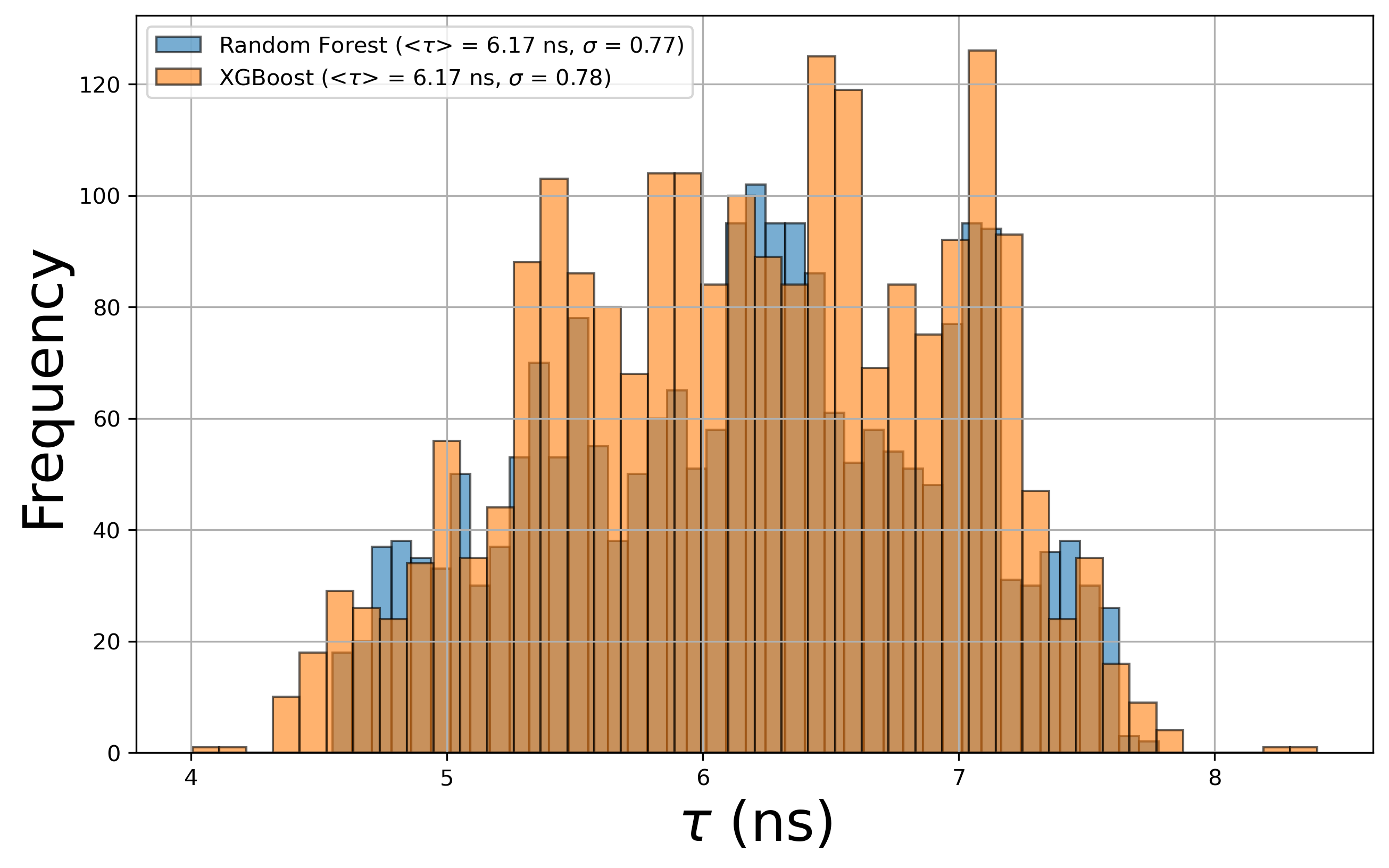}
\hfill
\caption{Left : \( 48 \times 48 = 2304 \) pixel map of the predicted dead time $\tau$ for the full Mini-EUSO focal surface, averaged over both ML models. Right : Comparison of the distribution of predicted dead times across all pixels for both models.}
\label{fig:minieusores}
\end{figure}

The results of the ML-based pile-up analysis are summarized in Figure~\ref{fig:minieusores}, which shows the averaged per-pixel values of $\tau$ predicted by both the Random Forest and XGBoost models.

With pixel-wise estimates of $\tau$ in hand, Eqs.~(\ref{eq:rho_from_N}) and~(\ref{eq:sigma_rho}) can be applied to recover the true photon fluxes across the focal surface and their uncertainties. The uncertainty on the estimated $\tau$, $\Delta\tau$, can be estimated and injected in Eq.~(\ref{eq:sigma_rho}). This refined unpile-up approach enhances the precision of flux reconstruction, supporting more accurate analysis of the Mini-EUSO data, including meteors, transient luminous events, or any other type of events.

\section{Conclusion and Outlook}

We presented a comprehensive framework for modeling, simulating, and correcting pile-up effects in event counting systems operating under extended dead time. Both this framework and experimental measurements showed excellent agreement, validating the extended dead-time model and enabling accurate event rate reconstruction even in high-rate regimes.

This framework was applied to Mini-EUSO data, where supervised learning techniques were used to infer per-pixel dead times from long-term histograms. The resulting dead time map allows for refined pile-up corrections across the focal surface, improving flux recovery for transient events such as ELVES and meteors.

This approach will be systematically applied to photodetectors in future JEM-EUSO missions, including the balloon-borne PBR project, where both pre-flight and in-flight pile-up analyses will be performed. More broadly, the framework is applicable to any counting system affected by dead time, with relevance to instrumentation in astrophysics or other fields.

\bibliographystyle{unsrt}
\bibliography{references}

@article{pomme2015,
  author    = {S. Pommé and others},
  title     = {Uncertainty of nuclear counting},
  journal   = {Metrologia},
  volume    = {52},
  pages     = {S3},
  year      = {2015},
  doi       = {10.1088/0026-1394/52/3/S3}
}

@techreport{muller1977,
  author    = {J. W. Müller},
  title     = {Asymptotic results for a modified renewal process and their application to counting distributions},
  institution = {BIPM},
  year      = {1977},
  number    = {77/1}
}

@techreport{muller1971,
  author    = {J. W. Müller},
  title     = {Interval densities for extended dead times},
  institution = {BIPM},
  year      = {1971},
  number    = {112}
}

@article{callibration1,
  author    = {E. Parizot and others},
  title     = {Characterization and absolute calibration of {R}-11265 multi-anode photomultiplier tubes for the {JEM-EUSO} space and balloon program: {I}. Methods and generic features},
  journal   = {Astroparticle Physics},
  volume    = {171},
  year      = {2025},
  doi       = {10.1016/j.astropartphys.2025.103112}
}

@article{callibration2,
  author    = {D. Trofimov and others},
  title     = {Characterization and absolute calibration of {R}-11265 multi-anode photomultiplier tubes for the {JEM-EUSO} space and balloon program: {II}. Application to the EUSO-SPB2 photodetection modules},
  journal   = {Astroparticle Physics},
  volume    = {172},
  year      = {2025},
  doi       = {10.1016/j.astropartphys.2025.103131}
}

@article{ASIC,
  author    = {S. Blin and others},
  title     = {{SPACIROC3}: 100 {MH}z photon counting {ASIC} for {EUSO-SPB}},
  journal   = {Nuclear Instruments and Methods in Physics Research Section {A}: Accelerators, Spectrometers, Detectors and Associated Equipment},
  volume    = {912},
  year      = {2018},
  doi       = {10.1016/j.nima.2017.12.060}
}

@article{Mini-EUSO,
  author    = {S. Bacholle and others},
  title     = {Mini-{EUSO} Mission to Study Earth {UV} Emissions on board the {ISS}},
  journal   = {ApJS},
  volume    = {253},
  pages     = {36},
  year      = {2021},
  doi       = {10.3847/1538-4365/abd93d}
}

    \newpage
{\Large\bf Full Authors list: The JEM-EUSO Collaboration}

\begin{sloppypar}
{\small \noindent
M.~Abdullahi$^{ep,er}$              
M.~Abrate$^{ek,el}$,                
J.H.~Adams Jr.$^{ld}$,              
D.~Allard$^{cb}$,                   
P.~Alldredge$^{ld}$,                
R.~Aloisio$^{ep,er}$,               
R.~Ammendola$^{ei}$,                
A.~Anastasio$^{ef}$,                
L.~Anchordoqui$^{le}$,              
V.~Andreoli$^{ek,el}$
% ...
}
\end{sloppypar}
\vspace*{.3cm}

{ \footnotesize
\noindent
%
% ...
$^{cb}$ Universit\'e de Paris, AstroParticule et Cosmologie (CNRS), Paris, France\\
% ...
$^{ef}$ Istituto Nazionale di Fisica Nucleare (INFN), Sezione di Napoli, Naples, Italy\\
% ...
$^{ei}$ Istituto Nazionale di Fisica Nucleare (INFN), Sezione di Roma Tor Vergata, Rome, Italy\\
% ...
$^{ek}$ Istituto Nazionale di Fisica Nucleare (INFN), Sezione di Torino, Turin, Italy\\
$^{el}$ Universit\`a di Torino, Dipartimento di Fisica, Turin, Italy\\
% ...
$^{ep}$ Gran Sasso Science Institute (GSSI), L’Aquila, Italy\\
$^{er}$ Istituto Nazionale di Fisica Nucleare (INFN), Laboratori Nazionali del Gran Sasso, Assergi, Italy\\
% ...
$^{ld}$ University of Alabama in Huntsville, Huntsville, AL, USA\\
$^{le}$ City University of New York (CUNY), Lehman College, Bronx, NY, USA\\
% ...
}

\end{document}